\documentclass[[english,british]{aa}
\usepackage[T1]{fontenc}
\usepackage{amsmath}
\usepackage{graphicx}
\usepackage[authoryear]{natbib}
\usepackage{epsfig}
\usepackage{txfonts}
\usepackage{color}
\usepackage{babel}
\providecommand{\tabularnewline}{\\}
\def\kms{km\,s$^{-1}$}

\def\vs{$v\sin{i}$}
\def\te{$T_{\rm eff}$}
\def\lg{$\log{g}$}

\def\te{$T_{\rm eff}$}

\def\kic{KIC\,4247791}

\begin{document}

\title{\kic: A SB4 system with two eclipsing binaries (2EBs)}

\subtitle{A quadruple system?
\thanks{Based on observations with the 2-m Alfred-Jensch-Telescope of the
Th\"uringer Landessternwarte Tautenburg.}}

\author{H. Lehmann\inst{1}\and M. Zechmeister\inst{2}\and S. Dreizler\inst{2}\and
S. Schuh\inst{2} \and R. Kanzler\inst{2}}

\institute{Th\"{u}ringer Landessternwarte Tautenburg (TLS), Sternwarte 5, 07778
Tautenburg, Germany\\ 
\email{lehm@tls-tautenburg.de}\and 
Georg-August-Universit\"{a}t, Institut f\"{u}r Astrophysik, Friedrich-Hund-Platz
1, 37077 G\"{o}ttingen, Germany\\
\email{zechmeister(dreizler,schuh,rkanzler)@astro.physik.uni-goettingen.de}}

\date{Received / Accepted}

\abstract{\kic\ is an eclipsing binary observed by the Kepler satellite mission.}
{We wish to determine the nature of its components and in particular the origin of a shallow dip in its Kepler light curve 
that previous investigations have been unable to explain in a unique way.}
{We analyse newly obtained high-resolution spectra of the star using synthetic spectra based on atmosphere models,
derive the radial velocities of the stellar components from cross-correlation with a synthetic template, and calculate the orbital solution.
We use the JKTEBOP program to model the Kepler light curve of \kic.}
{We find \kic\ to be a SB4 star. The radial velocity variations of its four components can be explained 
by two separate eclipsing binaries.  In contradiction to previous photometric findings, 
we show that the observed composite spectrum as well as the derived masses of all four of its components correspond to spectral type F.}
{The observed small dip in the light curve is not caused 
by a transit-like phenomenon but by the eclipses of the second binary system.
We find evidence that \kic\ might belong to the very rare hierarchical SB4 systems 
with two eclipsing binaries.}


\maketitle

\section{Introduction}
The Kepler mission was designed and launched with the primary aim
of searching for transiting planets, and indeed, has found hundreds
of planetary candidates \citep{Borucki2011ApJ...728..117B} and a large
number of eclipsing binaries (EBs). In addition to those already very interesting
objects, even more unusual systems such as multiple
transiting planets \citep{Lissauer2011arXiv1102.0543L} or triply
eclipsing triples \citep{Carter2011Sci...331..562C,Derekas2011Sci...332..216D,2011AJ....142..160S}
have been discovered.

\begin{figure}
\includegraphics[angle=-90, width=1.\linewidth]{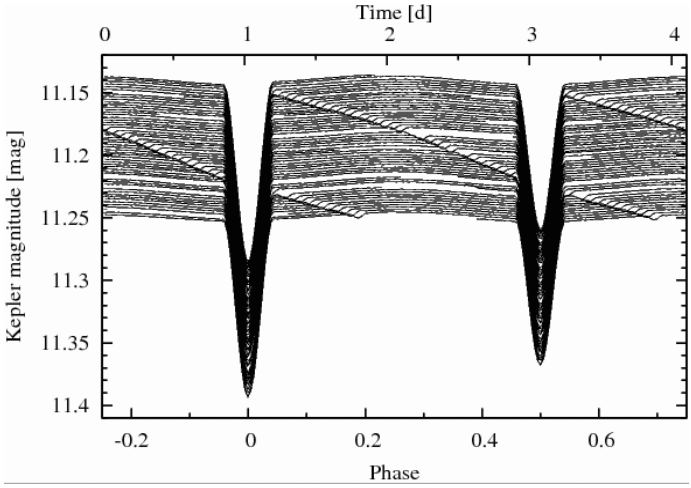}
\caption{\label{fig:light-curve}Detrended Kepler light curve (Q0-Q3 data) phase folded to 4\fd100871. 
A shallow dip drifts with time. Around 55 cycles are covered,
each shifted upwards by 2\,mmag (times the cycle number) for visibility.}
\end{figure}

In their analysis of low-mass EBs in the Kepler Q0 and Q1 data releases, \cite{Coughlin2011AJ....141...78C},
hereafter C11,
find that one object does not belong to the main-sequence, \object{KIC 4247791}\,=\,2MASS J\,19083956+3922369. 
The authors determine its temperature to be \te=\,4\,063\,K, an orbital period of
4\fd100866, a combined mass of 1.28\,M$_\odot$, and a combined radius of 3.82\,R$_\odot$.
According to these values, the system must contain at least one evolved component.
Moreover, a periodic transit-like feature is superimposed on the EB light curve. 
With respect to the main eclipse,
this feature drifts with time (Fig.~\ref{fig:light-curve}) and the authors derive possible periods of either 2\fd02484
or 4\fd04969. According to C11,
the feature could be explained in the case of the shorter period by (1) a background EB with no 
visible secondary, 
(2) a circumbinary transiting object, or (3) a transiting object around one of the stars in an almost 2:1 resonant 
orbit with the binary. In the case of the longer period, the explanation could be (4) a fore- or background EB 
with 0.98752 times the main period of \kic\ that has nearly identical depths of the primary and secondary eclipses.

 \kic\ also has an entry in the Kepler eclipsing binary catalogue \citep{2011AJ....141...83P} and both EBs
are included in its second release \citep{2011AJ....142..160S}.

We obtained 15 high-resolution spectra covering the orbital period of \kic\ well and use them
in this investigation to find the real cause of the observed transit-like feature in the Kepler light curve
of the star and to derive tighter constraints on the fundamental parameters of its components. For this purpose,
we measure the radial velocities (RVs) of the components of \kic, analyse the averaged composite spectrum, 
derive the orbits of the found binary systems, and combine the results with those obtained
from the Kepler light curve analysis.

\section{Observations}
All spectra have been taken with the Coude-Echelle spectrograph attached to the
2-m Alfred-Jensch-Telescope of the Th\"uringer Landessternwarte Tautenburg. The
spectra cover the wavelength range 4\,700\,\AA\ to 7\,400\,\AA\ with a resolution of
32\,000. The exposure time was 40 min per spectrum. 
The dates of observation are listed in Table\,\ref{Obs}.
Spectrum reduction was done using standard ESO-MIDAS packages including bias and stray-light subtraction,
filtering of cosmic rays, flat-fielding using a halogen lamp, optimal order extraction,
wavelength calibration using a ThAr lamp, and normalization to the local continuum. Nightly shifts in the instrumental zero
point were corrected by using a large number of telluric O$_2$ lines. 

\section{Spectrum analysis\label{sec:Spectrum-analysis}}

\begin{figure}
\includegraphics[angle=-90, width=1.\linewidth, clip=]{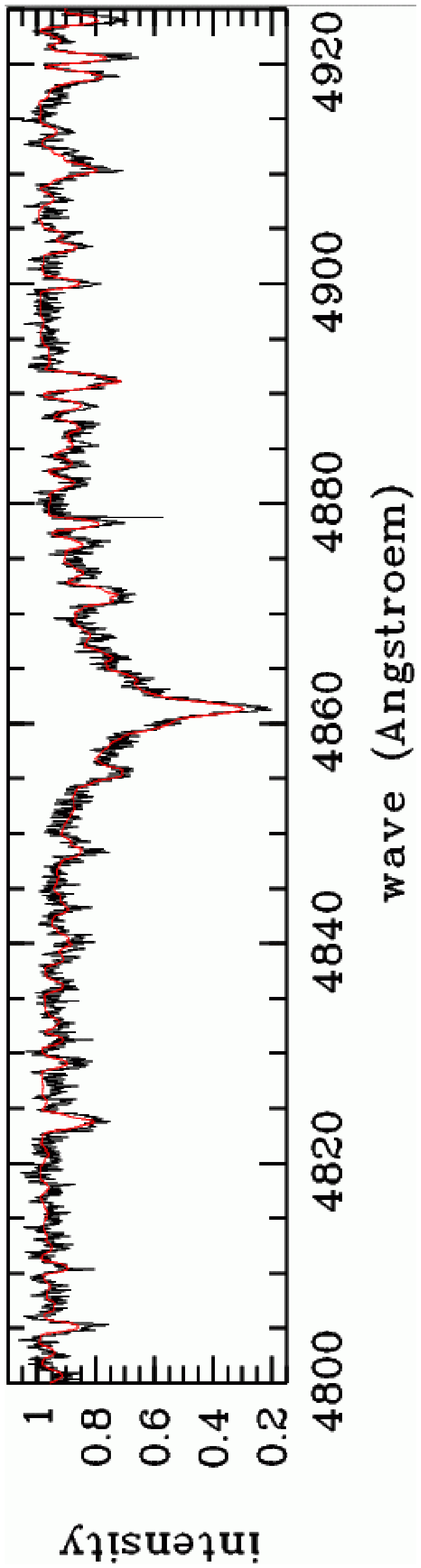}
\caption{Observed spectrum  in the H$\beta$ region (black, BJD= 2\,455\,819) and best-fit synthetic spectrum 
(red) computed for \te=6\,400\,K, \lg=3.7, [M/H]=$-0.2$\,dex, and \vs=40\,\kms.}
\label{Spec}
\end{figure}

In a first step, we determined the predominant spectral type from one observed, composite 
spectrum showing the sharpest (unsplit) lines. 
For the determination of the stellar parameters, we used a method identical to that described in
\cite{2011A&A...526A.124L}, to derive values of
\te=\,6\,400\,K, \lg\,=\,3.7, \vs\,=\,40\,\kms, and a metallicity of $-$0.2\,dex relative to the solar value.
Figure\,\ref{Spec} compares an observed spectrum with the best-fitting synthetic spectrum 
for the H\,$\beta$ region. The results are
representative of the composite spectrum. This spectrum has to be decomposed to derive the spectral types
of the single components. The derived temperature corresponds to spectral type F and contradicts the value of 4\,063\,K given by C11.
In particular, the observed Balmer line profiles are incompatible with such a low temperature value.

\section{Orbital solutions from cross-correlation}

\begin{figure}
\includegraphics[angle=-90, width=0.64\linewidth, clip=]{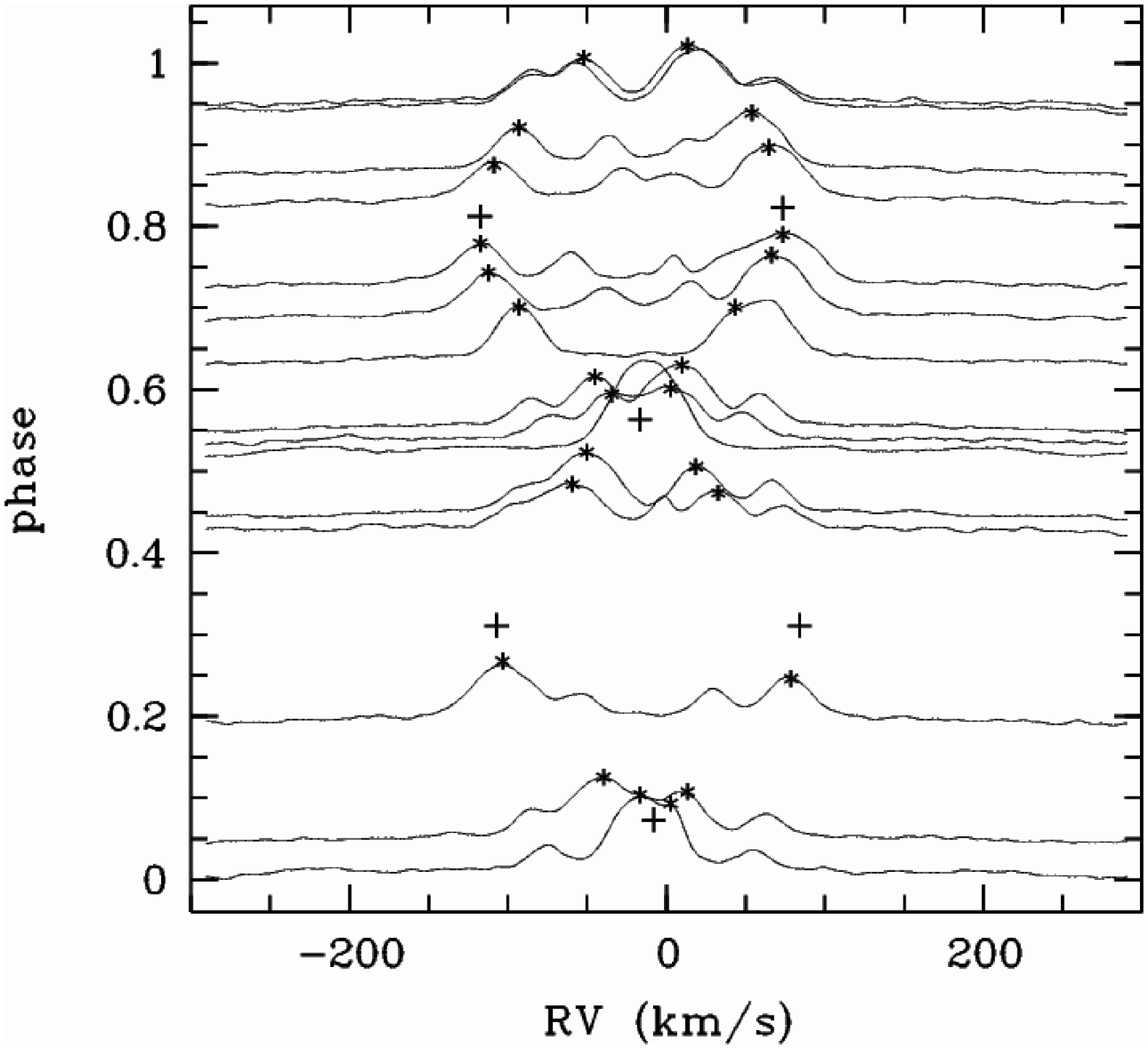}\vspace{-4mm}
\includegraphics[angle=-90, width=0.64\linewidth, clip=]{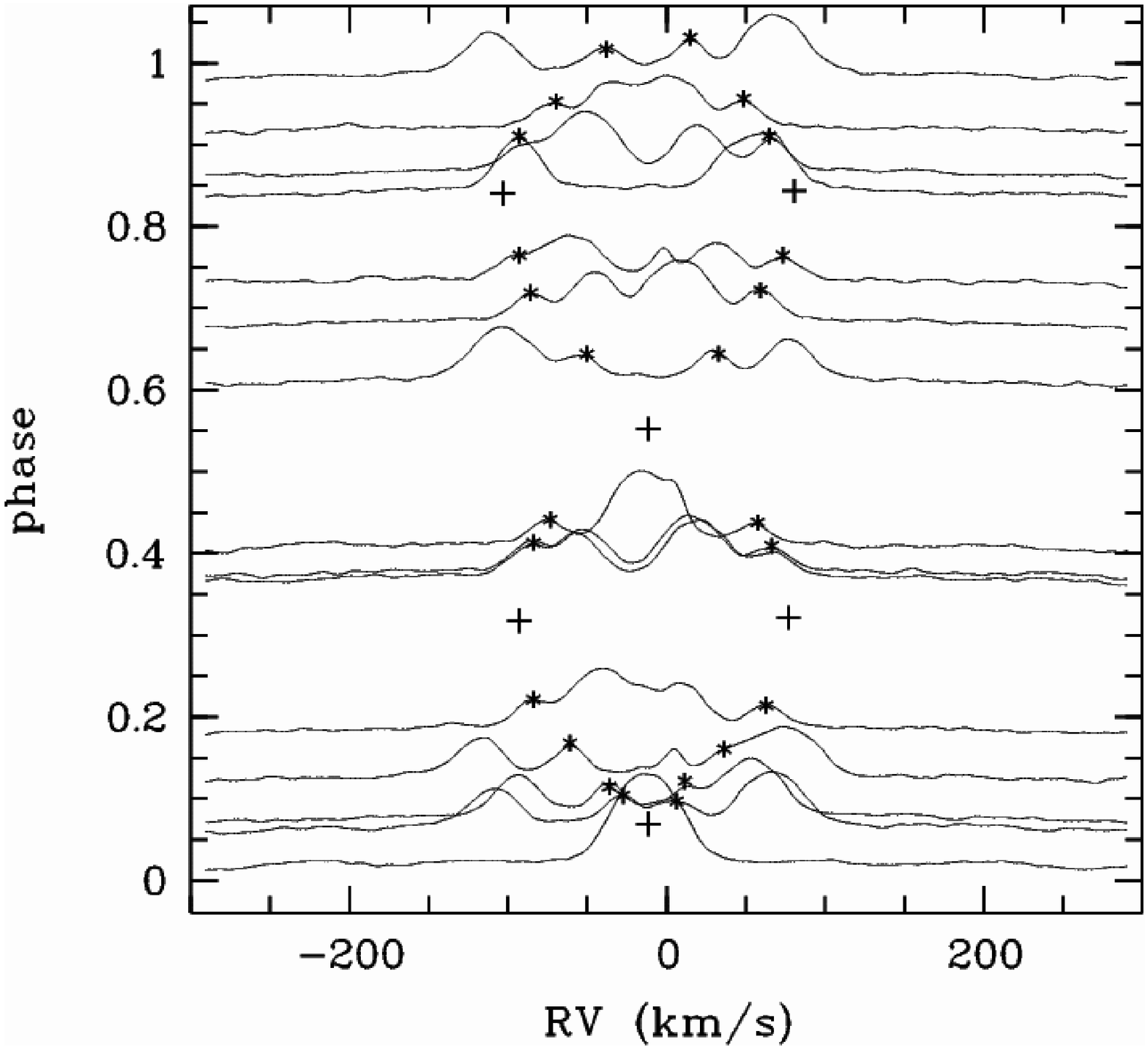}
\caption{CCFs vertically shifted with orbital phase based on periods of 4\fd100866 (top) and
4\fd049687 (bottom). The peak positions belonging to the corresponding orbit are marked by
asterisks, and the positions of minimum and maximum RV separation by plus signs.}
\label{CCFs}
\end{figure}

\begin{figure}
\includegraphics[angle=-90, width=0.85\linewidth, clip=]{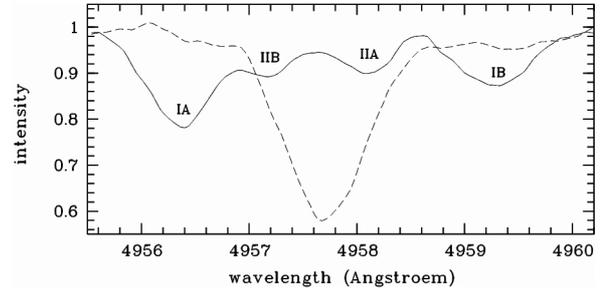}
\caption{The Fe\,I\,4\,957\,\AA\ line observed at two different orbital phases showing strong (continuous line) and
almost no (dashed line) splitting.}
\label{lines}
\end{figure}

Cross-correlation functions (CCFs) were obtained from the cross-correlation of the observed spectra with the
synthetic template spectrum derived in Sect.\,\ref{sec:Spectrum-analysis} but using 
\vs\,=\,1\,\kms\ instead of 40\,\kms\ to obtain the maximum spectral resolution in the CCFs. In most of the CCFs,
we discovered four components. The positions of
these components represent the RVs of four stars in two binary systems with orbital periods 
close to the periods of 4\fd100866 (the two stronger peaks in the CCFs) and
4\fd049687 (the two smaller peaks) given in C11. This can be seen clearly if we arrange the CCFs according to 
the positions in orbital phase based on the two periods (Fig.\,\ref{CCFs}).  

The line splitting can also be seen in the single spectra. Figure\,\ref{lines} compares the 
Fe\,I 4\,957\,\AA\ line from two spectra taken at different orbital phases where the noise was 
reduced by averaging over 9 wavelength bins. One spectrum clearly shows a splitting into four 
components belonging to system I, its stars A and B, and system II, its stars A and B, whereas the 
other spectrum has almost no splitting.

The RVs were determined from the CCFs using multiple Gaussian fits. From 14 spectra, we could 
determine the RVs of all components. We used four Gaussians in 12 cases and five Gaussians in the 2 cases where we
detected a fifth contribution produced by weak telluric lines. In one spectrum,
we could only resolve the RVs of the brighter EB. 
Table\,\ref{Obs} lists the RV values obtained for the four components.

\begin{figure}
\includegraphics[width=0.92\linewidth, clip=]{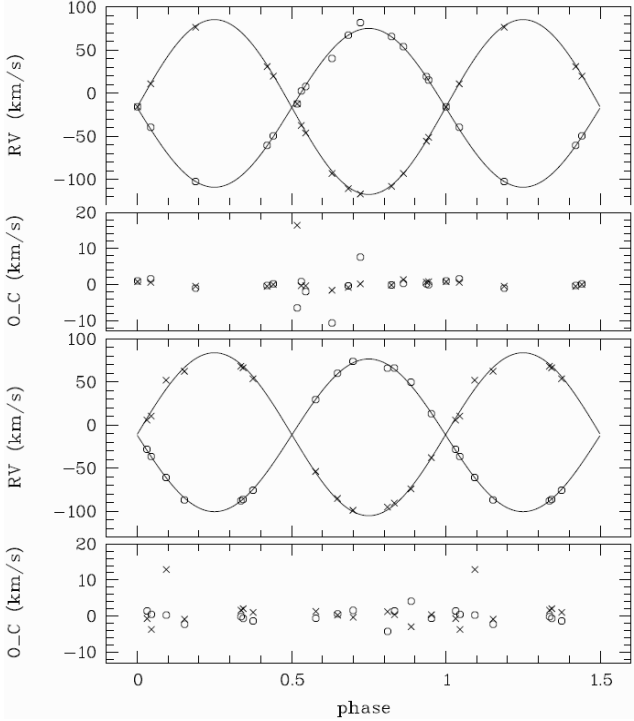}
\caption{From top to bottom: RVs of components IA (circles) and IB
(crosses) folded with the period of 4\fd100871, O$-$C residuals of IA and IB, RVs of components IIA 
(circles) and IIB (crosses) folded with the period of 4\fd049732, and O$-$C residuals of IIA and IIB.}
\label{orbits}
\end{figure}

The orbital solutions were calculated using the method of differential corrections to the orbital elements 
\citep{1941PNAS...27..175S}. We assumed circular orbits in each case. For both systems, we fixed the orbital 
periods and times of $\gamma$-velocity passage/primary minimum (iterated with those obtained in Sect.\,\ref{lca}). 
Figure\,\ref{orbits} shows the RVs of both systems and the corresponding O$-$C values folded with the two different periods.
Three outliers can be seen for component IA, one for IB and one for IIB. These outliers were rejected when calculating the
orbital solutions. They can arise when the RV components coincide and are poorly resolved 
in the CCFs, in particular at/around phases 0 and 0.5, which moreover can be affected 
by the Rossiter effect (for A and B, respectively).
The derived orbital elements are listed in the upper part of Table\,\ref{tab:elements}.

Phase zero in Fig.\,\ref{orbits} corresponds to the time of primary minimum ($T_{\mathrm{MIN}}$ in Table\,\ref{tab:elements}).
We see that in both systems the RVs of the more massive components A that have slightly 
smaller RV amplitudes, decrease at this phase. This means that the more massive components
are eclipsed at $T_{\mathrm{MIN}}$ and should have higher surface brightnesses.
It is remarkable that not only the periods of the two systems are almost identical (but significantly different
as can also be seen from their Kepler light curves) but that also the RV semi-amplitudes $K$ and the mass ratios $q$ are very similar.

\begin{table}
\caption{\label{tab:elements}Parameters and spectral types (SpT) derived from the spectroscopic, light curve, and combined analysis. 
$T_{\mathrm{MIN}}$ gives BJD--2\,455\,000, errors in units of the last digits are given in parentheses.}
\tabcolsep 1.77mm
\begin{tabular}{|l|ll|ll|} 
\hline  &   \multicolumn{2}{c|}{system I}  & \multicolumn{2}{c|}{system II}\tabularnewline
 &  \multicolumn{1}{c}{A}  & \multicolumn{1}{c|}{B} & \multicolumn{1}{c}{A}  & \multicolumn{1}{c|}{B} \tabularnewline
\hline
$K$                 [\kms]	     & ~~~91.9(4) & ~101.6(3) & ~~~88.6(3) & ~~~91.4(4)     \tabularnewline
$\gamma$            [\kms]	     & $-16.2(9)$ & $-17.1(8)$ & $-11.8(2.0)$ & $-11.7(1.7)$	 \tabularnewline
$q$                		     & \multicolumn{2}{c|}{0.905(7)} & \multicolumn{2}{c|}{0.970(7)}	 \tabularnewline
\hline
$P$                 [d] 	     & \multicolumn{2}{c|}{4.100871(4)} & \multicolumn{2}{c|}{4.049732(71)}	\tabularnewline
$T_{\mathrm{MIN}}$  [BJD]	     & \multicolumn{2}{c|}{~~~~1.1453(1)} & \multicolumn{2}{c|}{~~~~~~~~1.353(1)}     \tabularnewline
$i$                 [deg]	     & \multicolumn{2}{c|}{~~~~~~~79.5(1)} & \multicolumn{2}{c|}{~~~~~~~~~~80.0(8)}	\tabularnewline
$e\cos\omega$      		     & \multicolumn{2}{c|}{~0.00012(6)} & \multicolumn{2}{c|}{~~~~~~0.0020(3)}     \tabularnewline
$e\sin\omega$      		     & \multicolumn{2}{c|}{-0.00570(4)} & \multicolumn{2}{c|}{~~~~~~0.0009(4)}     \tabularnewline
$e$                		     & \multicolumn{2}{c|}{~0.00570(4)} & \multicolumn{2}{c|}{~~~~~~0.0022(3)}     \tabularnewline
$\omega$            [deg]	     & \multicolumn{2}{c|}{~~~178.7(6)} & \multicolumn{2}{c|}{~~~~~~~~~~~65(24)}     \tabularnewline
$R/a$              		     & ~~~0.1570(4) & ~~~0.1503(4) & ~~~0.100(9) & ~~~0.092(8)     \tabularnewline
$L/L_{\mathrm{AB}}$		   & ~~~0.576(1) & ~~~0.424(1) & ~~~0.55(1) & ~~~0.45(1)     \tabularnewline
$L_{\mathrm{AB}}/L_{\mathrm{I+II}}$  & \multicolumn{2}{c|}{~~~~~0.76(2)} & \multicolumn{2}{c|}{~~~~~~~~~0.24(1)}     \tabularnewline
\hline
$a$                [AU] 	    & \multicolumn{2}{c|}{~0.0742(2)} & \multicolumn{2}{c|}{~~~~~0.0681(2)}	\tabularnewline
$R$                [R$_{\odot}$]    & ~~~2.50(1) & ~~~2.40(1) & ~~~1.46(13) & ~~~1.35(12)     \tabularnewline
$M_{\mathrm{AB}}$  [M$_{\odot}$]    & \multicolumn{2}{c|}{~~~~3.24(3)} & \multicolumn{2}{c|}{~~~~~~~2.56(3)}	 \tabularnewline
$M$                [M$_{\odot}$]    & ~~~1.70(2) & ~~~1.54(1) & ~~~1.30(2) & ~~~1.26(2)     \tabularnewline
$\log g$           [dex]	    & ~~~3.87(1) & ~~~3.87(1) & ~~~4.22(8) & ~~~4.28(8)     \tabularnewline
SpT               		    & ~~~~F0\,IV & ~~~~F2\,IV & ~~~~F7\,V & ~~~~F8\,V	\tabularnewline
\hline
\end{tabular}
\end{table}

\section{Light curve analysis\label{lca}}

The Kepler database provides publicly available data for 228 days (2009-05-02 to 2009-12-16; quarters Q0-Q3) of 
photometric observations of \kic\ (Fig.\,1), including 88 days taken in short cadence mode 
(SC, Q2, 1 min exposure time).
On the basis of JKTEBOP
\citep{Southworth2004MNRAS.351.1277S,Southworth2004MNRAS.355..986S}, the
light curve was analysed {\em simultaneously} for both EBs using the IDL
implementation {\tt mpfit} of the  Levenberg-Marquardt algorithm.
 A fit was made to the two binary light curves using 14 free parameters in each case: 
binary period, phase zero-point, ratio and sum of the radii, inclination, 
surface brightness ratio, eccentricity ($e\sin\omega, e\cos\omega$), 
quadratic limb-darkening, and reflection effect coefficients. 
The mass ratio was fixed to the values
determined from the RV measurements. The relative
brightness ratio as well as a contribution from a (constant) fifth
background light was also taken into account. 

The Kepler light
curves are influenced by instrumental effects, which were treated separately
for each quarter or even for sections separated by an interruption of
the observations. This is fitted simultaneously with the binaries by
piecewise second or third order polynomials.
Starting from an initial guess for a total of 52 parameters using
identical stellar parameters, zero eccentricity, zero fifth light, and neglecting systematic trends, 
we employed the IDL routine {\tt mpfit} to perform an optimisation. {\tt mpfit} calls 
JKTEBOP twice to evaluate the contribution of each
binary. Since several of the parameters are closely related, the
result of
the optimisation depends partly on the initial conditions. After a
5-$\sigma$ clipping to remove outliers, we therefore started 50
optimisation runs where the initial conditions were randomly varied
around the first result. The one with the lowest $\chi2$ was taken as
the final result. {\tt mpfit} also provides errors from the diagonal
of the covariance matrix. For more realistic errors in the presence of
correlated noise, we however, followed
the approach of \cite{2008MNRAS.386.1644S}. To take non-Gaussian residuals into
account, a series of 500 synthetic light curves was constructed using
the best-fit model adding arbitrarily shifted residuals as noise. Since
the residuals from the long and short cadence data differed
significantly, they were treated separately. These light curves
were again fitted using various start values. From the nearly Gaussian
distribution of the parameters around the best-fit relation, we determined the\linebreak
1-$\sigma$ uncertainty limits. These are typically an order of
magnitude larger than the formal errors provided by {\tt mpfit}

\begin{figure}
\includegraphics[angle=-90, width=1\linewidth, clip=]{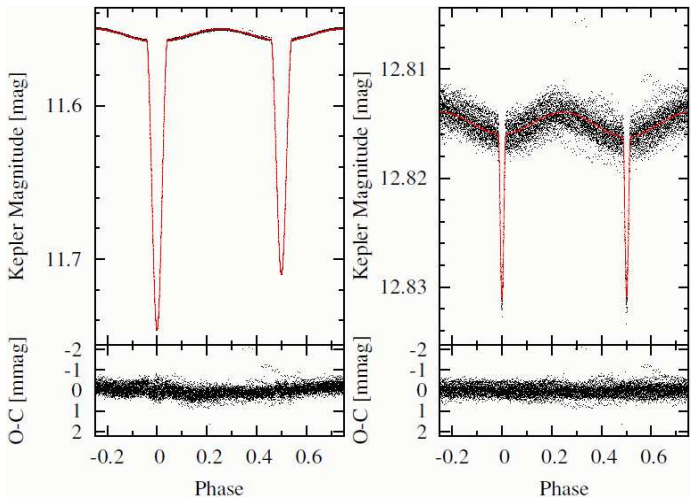}
\caption{Disentangled and phase-folded light curves. Left: Light curve (EB\,II model subtracted) fitted by
the EB\,I model. Right: Light curve (EB\,I model subtracted) fitted by the EB\,II model. The 
corresponding O$-C$ residuals are shown in the lower panels.}
\label{phased}
\end{figure}

 The derived parameters are listed in Table\,\ref{tab:elements}, and Fig.\,\ref{phased} 
illustrates the quality of the fit.
In addition, we verified the results by fitting each EB separately. For this purpose, the data were decomposed 
iteratively by phase folding the light curve to the periods of the EBs and calculating the medians in one-minute 
bins. Many cycles had to be be covered to minimise the residual correlation effects introduced 
by the ellipsoidal variations and the similar periods of both EBs. The Q2-SC data covers 
about 21 orbit cycles of EB\,I and 27\% of the full beat cycle (324.5 days) of both EBs.

With the known periods, inclinations, and RV amplitudes of each component,
we were able to derive the stellar radii and masses listed in Table\,\ref{tab:elements}.  
From the masses and radii, we roughly estimated the spectral types using the
tables of \citet{Gray2005}. We derived a spectral type F for all four stars, which agrees
with the observed composite spectrum.
For main-sequence stars,
we would expect radii of about 1.5\,R$_\odot$ for the stars of system I and
of 1.3\,R$_\odot$ for the stars of system II and \lg\ of about 4.3 for all the components \citep{Gray2005}.
Comparing these values with Table\,\ref{tab:elements}, we see that both components of system\,I 
must be slightly evolved.

From the derived orbital period and the stellar radii, we were able to estimate the \vs\ of the components
of system\,I assuming synchronized rotation. 
We measured 31\,\kms\ for component A and 30\,\kms\ for B. This is compatible with the 40\,\kms\
derived from the composite spectrum in Sect.\,\ref{sec:Spectrum-analysis} that we should recall is affected 
by an additional broadening since this spectrum was taken close to but not exactly during 
the (double) $\gamma$-velocity passage.

\section{Discussion}

\begin{figure}
\includegraphics[angle=-90, width=1.\linewidth, clip=]{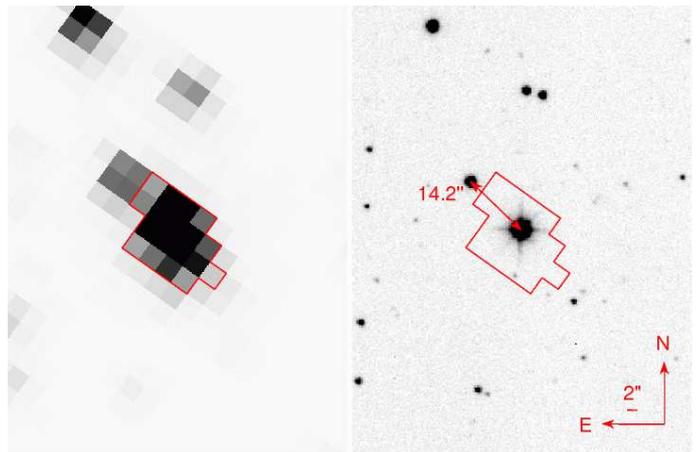}
\caption{Left: 1.12\arcmin x 1.50\arcmin section of a Kepler full-frame image (FFI). The used extraction aperture 
for \kic\ is marked 
by the red borders. Right: UKIRT $J$-band image of the same region. The Kepler aperture is overlaid. The marked, nearby 
blend is KIC\,4247807.}
\label{Pixel}
\end{figure}

Our results clearly show that the effects observed in the Kepler light curve
can be explained by two EBs and not by some transit-like phenomenon.
It is difficult to understand, however, that all the similarities in the spectroscopically derived
parameters of the two binary systems are caused by chance and one has to ask 
wether there is a possible connection between the systems.
Therefore, the most interesting question is whether both EBs are gravitationally
bound and form a quadruple system. 

 To exclude visible blends, we did a careful search for any contaminants in the Kepler pixel data and in
available ground-based images of the corresponding field of view. 
\kic\ can be found in one J-band image of the UKIRT InfraRed Deep Sky Survey 
\citep[UKIDSS\footnote{http://surveys.roe.ac.uk/wsa}, DR\,WSERV4, see][]{2007MNRAS.379.1599L},
as well as in the Sloan Digital Sky Survey (SDSS-III\footnote{http://www.sdss3.org}, DR8) 
in all five ($u$, $g$, $r$, $i$, $z$) bands. The UKIRT image has the higher spatial resolution with a pixel 
scale of 0\farcs2/pixel but the image of \kic\ itself is overexposed. The SDSS provides data of 0\farcs396/pixel
and \kic\ is non-saturated in $u$ and $z$, where the Gunn-z image has the better seeing.
We used the UKIRT $J$-band image to check for possible blends in the vicinity of the star and the SDSS $z$-band image
to investigate its flux profile.

An inspection of the $J$-band image and a comparison with the pixel data of the Kepler images identified an 
object (\object{KIC 4247807}, Kepler magnitude $K_p$\,=\,14.808\,mag) separated by 
14\farcs2 from \kic\ ($K_p$\,=\,11.260\,mag) close to the aperture used for extraction (Fig.\,\ref{Pixel}). 
However, the following finding excludes the possibility that KIC 4247807 is EB\,II: the spectra were taken 
with a 2\arcsec\ slit, where the slit orientation on the sky varies with the hour angle of the telescope. For any
object that did not lie completely inside the slit, we were unable to derive as precise a RV curve as we obtained from the
spectral features of EB\,II. 
The close object is 3.5 mag fainter than \kic. Assuming that 10\% of the light of KIC 4247807 
from the PSF wings falls into the aperture, its fifth light contribution should be at a level well below
1\%.

\begin{figure}
\includegraphics[angle=-90, width=1\linewidth, clip=]{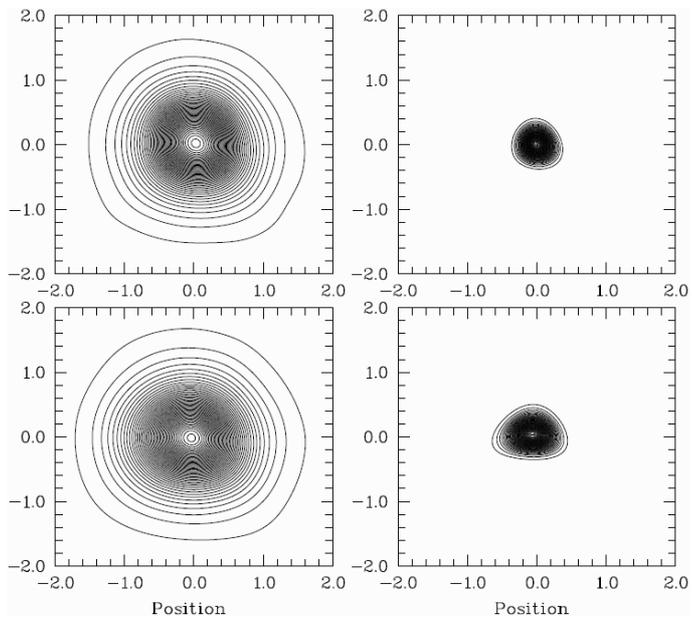}
\caption{Top: Contour plots of the SDSS $z$-band image of \kic\ (left) and of the image deconvolved with the PSF (right).
The position scale is in arcsec.
Bottom: The same with a simulated faint companion in 0\farcs5 distance.}
\label{Contour}
\end{figure}

 Figure\,\ref{Contour} shows in its upper left panel a contour plot of the $z$-band image of \kic.
To enhance the resolution, we constructed the mean point spread function (PSF) from all brighter and non-saturated
star images in the $z$-band field by applying a shift and add, sampling all images to ten times higher resolution than provided
by the original ones. Two-dimensional Gaussian fits where used to eliminate all subimages showing any
asymmetry in the flux profiles. We ended up with a PSF built from 63 star images with a 1\farcs08 FWHM, the same 
value as we obtained from the image of \kic\ itself. The upper right 
panel in Fig.\,\ref{Contour} shows the image of the star deconvolved with the PSF using the Lucy-Richardson algorithm
\citep{1974AJ.....79..745L}. 
It is almost perfectly symmetric, possessing a FWHM of 0\farcs32.
For comparison, we simulated a nearby star of 0.2 times the brightness of \kic\ at a distance of 0\farcs5
using the derived PSF. The result is shown in the lower panels of Fig.\,\ref{Contour}. The deconvolved image
is clearly asymmetric having a FWHM of 0\farcs44 in the direction towards the faint companion and of 0\farcs35
in the perpendicular direction. 
We conclude that the separation between EB\,I and EB\,II must be closer than 0\farcs5,
which still implies that they may be either affected by an accidental blend or be gravitationally bound systems.

There are some findings that are consistent with a quadruple configuration of \kic\ such as the luminosity ratio 
that provides similar 
distances of both EBs or the similar inclinations and masses that are indicative of a common formation. 
The similarity between the systemic RVs also implies that it is a physical associated
system. The small difference in the system RVs of \textasciitilde{}5\,\kms\
could, however, be due to the orbital motion of both EBs around
each other and might be used to derive an upper limit to the separation
between both EBs. For a circular and edge-on
orbit, this upper limit would be 1200 years. We note that this limit is proportional
to $\sim$$K^{-3}$ meaning that if we observe e.g. only a quarter
of the maximum RV difference the period would shorten to about 20 years.
In the latter case, the corresponding variations in the systemic $\gamma$-velocities of the two systems
should be safely detectable from observations spanning about two years.

\section{Conclusions}

Using the period derived in C11 from the minima in the Kepler light curve,
we have been able to model the RV variations of the two main components identified in the CCFs derived from our newly obtained
spectra. Two additional features have been observed in almost all the CCFs, which vary with the period of 
4\fd049751 that was proposed by C11 as a possible period to explain the occurrence of a shallow dip in the 
Kepler light curve drifting with time against the main minima.

Our spectroscopic analysis has uncovered four stars moving in two separate binary systems.
The basic evidence for such an interpretation comes from the time-dependent
splitting of the spectral lines into four components that we have been able to assign
in a unique way to four stellar components moving in two separate orbits owing to
the small difference between the two periods being
significant, and visible as small moving dips in the Kepler light curve.

Our results for the combined photometric and spectroscopic observations are consistent with the scenario of two
binaries, both lying along the same line of sight, showing eclipses, having almost the same orbital periods,
RV amplitudes, and mass ratios. This seems to be a very rare if not unlikely
scenario and raises the questions of whether these two systems are physically connected.
The observed similarities may be indicative of a quadruple system, although we cannot
entirely rule out a moderate back- or foreground scenario.

 All derived fundamental parameters of EB\,I are well defined. For the primary and secondary, respectively,
we obtained masses of 1.70$\pm$0.02\,M$_\odot$ and 1.54$\pm$0.01\,M$_\odot$, and radii of 
2.50$\pm$0.01\,R$_\odot$ and 2.40$\pm$0.01\,R$_\odot$
and \lg\ of 3.87$\pm$0.01 for both components, corresponding to slightly evolved stars of early F-type. 
For EB\,II, the derived masses have similar accuracy to that of EB\,I but the errors in the radii and \lg\ 
are distinctly larger. We obtained
$M_{A}$=1.30$\pm$0.02\,M$_\odot$, $R_{A}$=1.46$\pm$0.13\,R$_\odot$, and \lg$_{A}$=4.22$\pm$0.08 for the primary and
$M_{B}$=1.26$\pm$0.02\,M$_\odot$, $R_{B}$=1.35$\pm$0.12\,R$_\odot$, and \lg$_{B}$=4.28$\pm$0.08\,M$_\odot$ 
for the secondary corresponding to late F-type main sequence stars.

To our knowledge, only two stars have been found so far to host two EBs in hierarchical systems, the visual 
binaries ADS\,9537\,AB and V994\,Her. ADS\,9537\,AB was discovered by
\citet{1965AJ.....70..666B} to consist of two W\,UMa-type EBs where the two systems orbit each other with a period 
of about 22\,000\,yr \citep{1986PASP...98...92B}. \citet{2008MNRAS.389.1630L} discovered that the SB4 star V994 Her 
hosts two EBs consisting of B8+A0 and A2+A4 main sequence stars. From astrometric data, they estimate a period of
the visual binary of few thousand years. There are a few other remarkable objects.
\citet{2008ApJ...682.1248S} discovered a SB4 quadruple
system, BD-22\,5866, consisting of one (K7+K7) EB and a second (M1+M2), non-EB system, orbiting each other with a period
shorter than 10 yr.
\object{OGLE 051343.14-691837.1} was proposed as a SB4 candidate with two EBs by \citet{2008IBVS.5868....1O} 
but has since been found to be SB2 
\citep{2010ASPC..435..403K, 2011IAUS..272..541R}. The SB3 quadruple system 
\object{V379 Cep} was found by \citet{2007A&A...463.1061H} to host two binaries where one is an EB.

Further observations would be necessary to confirm the quadruple nature of \kic, namely data 
 extending the present time base 
and designed to detect the orbital motion between systems
I and II based on transit timing or RV variations.
In a forthcoming paper, we intend to decompose the spectra of the 
components using the KOREL program \citep{Hadrava2006Ap&SS.304..337H}
and we will analyse in greater detail the decomposed light curves and spectra to 
provide more stringent constraints on the fundamental stellar parameters, in particular the effective
temperatures of the components,
and hence the evolutionary state and distance modulus of \kic.

\acknowledgements{This research has made use of SDSS-III.
Funding for SDSS-III has been provided by the Alfred P. Sloan Foundation, 
the Participating Institutions, the National Science Foundation, and the U.S. Department of 
Energy Office of Science. The SDSS-III web site is http://www.sdss3.org/.}

\bibliographystyle{aa}
\bibliography{AA_2011_18298_corrected}

\begin{appendix} 

\section{Measured radial velocities}
\begin{table}[h]
\tabcolsep 3.42mm
\caption{RVs in \kms\ derived from the four components in the CCFs where 
$\triangle RV$ gives the mean measurement error.}
\begin{tabular}{crrrr}
\hline\hline
BJD$-$2\,455\,000 & $RV_\mathrm{IA}$ & $RV_\mathrm{IB}$ & $RV_\mathrm{IIA}$ & $RV_\mathrm{IIB}$ \tabularnewline
\hline
700.529717  &	7.82 &  -46.25 & 60.25 &  -85.15 \tabularnewline
702.571156  & -39.50 &   10.92 &  -86.85 &   62.50 \tabularnewline
725.484044  &  40.41 &  -93.13 & 66.07 &  -95.22 \tabularnewline
726.433284  &  53.99 &  -93.06 &  -36.43 &   10.44 \tabularnewline
734.477064  &  65.83 & -108.09 &  -28.07 &    5.81 \tabularnewline
754.407123  &  67.44 & -110.81 & 13.18 &  -37.77 \tabularnewline
757.429167  & -60.46 &   30.93 & 73.98 &  -98.91 \tabularnewline
782.488513  &	2.61 &  -37.39 & 49.85 &  -74.03 \tabularnewline
787.374417  &  81.94 & -117.03 &  -60.90 &   52.13 \tabularnewline
788.514326  & -15.77 &  -15.77 &  -75.66 &   53.74 \tabularnewline
793.387013  & -102.52 &   76.62 & 29.65 &  -53.81 \tabularnewline
794.419142  & -49.49 &   19.83 & 66.29 &  -90.44 \tabularnewline
796.454312  &  19.35 &  -55.60 &  -87.92 &   68.29 \tabularnewline
796.483258  &  15.31 &  -51.26 &  -86.30 &   66.50 \tabularnewline
819.341840  & -12.25 &  -12.25 & --	&   --      \tabularnewline   
\hline
$\triangle RV$ & 0.15 & 0.15 & 0.28 & 0.24\tabularnewline
\hline
\end{tabular}
\label{Obs}
\end{table}

\end{appendix}

\end{document}